# Rate-potential decoupling: a biophysical perspective of electrocatalysis

Peter Agbo[1,2,3]*

[1]*Chemical Sciences Division*, [2]*Liquid Sunlight Alliance*, [3]*Molecular Biophysics & Integrated Bioimaging Division, Lawrence Berkeley National Laboratory, Berkeley, CA 94720, USA*

**Abstract**
In this perspective, the chemical physics of biological electron transfer are considered in relation to artificial electrocatalyst development. Nature's ability to access a wide range of chemical reactivities through a narrow set of redox-active motifs, in part by decoupling electron transport rates from reaction driving forces, is suggested as a model for the future of electrocatalyst design and testing. Theoretical rationale and experimental precedents for this concept are put forth, outlining how emulating nature's ability to arbitrarily tune tunneling currents with respect to donor/acceptor redox potentials – reaction driving forces – may enhance our control of electrocatalyst selectivity.

**Canonical bioinorganic catalysis: the assumed centrality of structure-function relationships**
Organometallic mimics of enzyme active sites, such as those of cytochrome P450 porphyrin centers and the oxygen evolving complex (OEC) of photosystem II, have substantially advanced our understanding of their underlying chemical mechanisms and spectroscopic features[1–3]. However, functional mimicry of redox enzymes, particularly through the use of organometallic complexes, has proven uniquely challenging to the field of bioinorganic chemistry. Attempts to link inorganic catalysis and enzyme biochemistry have generally focused on the development of homogeneous molecular catalysts, which act as structural mimics of biological active sites but typically ignore the higher-order aspects of protein design, particularly the roles played by secondary and tertiary structure, solvent exclusion effects and regulated electron transport in enzyme catalysis[4,5]. By and large, bioinorganic chemists have operated under an implicit assumption that structural mimicry of enzyme active sites should correspond to functional emulation of enzyme catalysis. However, this first approximation has been found to be largely insufficient, with molecular catalysts that superficially resemble protein active centers often failing to fully replicate enzyme reactivities.

While structure plays a critical role in facilitating enzyme catalysis, this clearly marks only one necessary aspect of functional biomicry. For example, while detailed molecular mimics of photosystem II's OEC have generally failed to fully reproduce the water oxidation chemistry it catalyzes, purely inorganic Mn oxide powders have been found to drive efficient water oxidation[6]. In the specific case of calcium-containing birnessite-like Mn oxides, these materials contain $Mn_3O_4Ca$ domains similar to the $CaMn_4O_5$ cubanes constituting the OEC, despite omitting the surrounding ligand framework present in both the OEC and organometallic model complexes[1,7,6]. The comparative success of inorganic powders (bearing little resemblance to protein active sites) relative to molecular catalyst active site mimics, suggest that structural emulation is not necessarily the key to achieving functional mimcry of enzymes. Instead, mastering enzymatic biomimicry at a fundamental, molecular level will ultimately require that the catalysts being developed – whether homogeneous or heterogeneous – are able to reproduce the underlying physics of the system we are seeking to replicate. In proteins, this requirement encapsulates a broad set of biophysical design components, including the regulation of charge transfer kinetics, the focus of this paper. In particular, this article focuses on the concept of decoupling current density (J) and potential (V) in electrocatalysis as a method for independently controlling charge transport kinetics and energetics in electrocatalysis. Exploration of this subject is motivated by the potential for increased control over carrier dynamics to facilitate more sophisticated characterization of artificial electrocatalysts, while introducing routes for improved control over

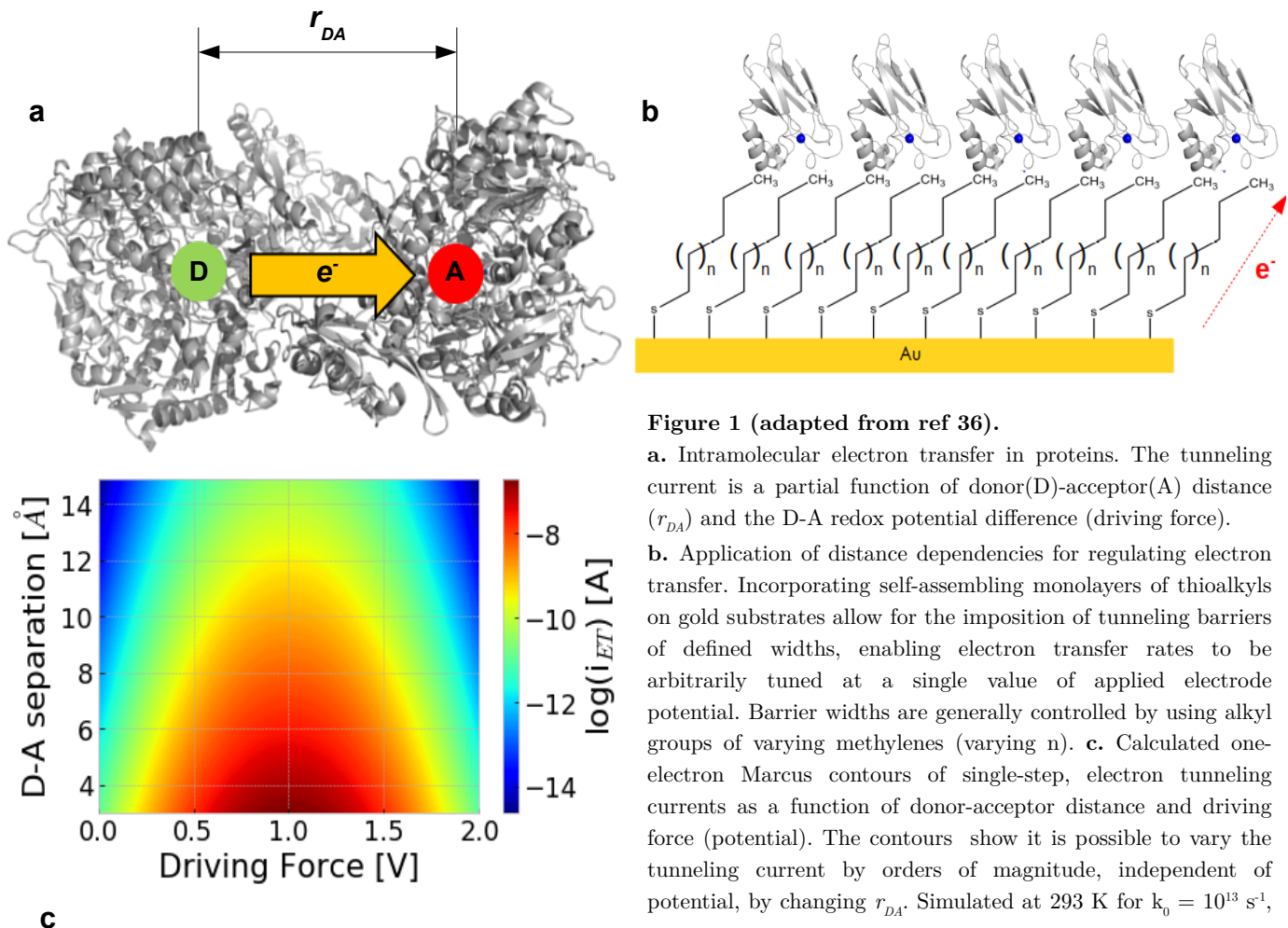

**Figure 1 (adapted from ref 36).**
**a.** Intramolecular electron transfer in proteins. The tunneling current is a partial function of donor(D)-acceptor(A) distance ($r_{DA}$) and the D-A redox potential difference (driving force).
**b.** Application of distance dependencies for regulating electron transfer. Incorporating self-assembling monolayers of thioalkyls on gold substrates allow for the imposition of tunneling barriers of defined widths, enabling electron transfer rates to be arbitrarily tuned at a single value of applied electrode potential. Barrier widths are generally controlled by using alkyl groups of varying methylenes (varying n). **c.** Calculated one-electron Marcus contours of single-step, electron tunneling currents as a function of donor-acceptor distance and driving force (potential). The contours show it is possible to vary the tunneling current by orders of magnitude, independent of potential, by changing $r_{DA}$. Simulated at 293 K for $k_0 = 10^{13}$ s$^{-1}$, $\lambda = 1.0$ eV, $\beta = 1.4$ Å$^{-1}$, calculated using eq. 1.

product selectivities in photoelectrochemical (PEC) systems.

**Current-potential decoupling and its correspondence with biological electron transfer**

The physical possibility of electrochemically decoupling current and potential is suggested by considering electron transfer in biochemical systems, where the decoupling of kinetics and energetics is indeed a physical reality for electronic carriers. In enzymes, the difference in redox potential ($\Delta \varepsilon^0$) between a donor (D) and acceptor (A) site dictates the magnitude of the electrical potential through which an electron is transferred. Meanwhile, the kinetics of macromolecular electron tunneling are determined by semi-classical Marcus Theory, an application of Fermi's Golden Rule where the rate of charge transfer ($dq/dt$) between a donor and acceptor is controlled by a number of factors, including the redox potential difference (driving force) between these sites *and* the distance ($r_{DA}$) between them[5,8–15]:

$$\frac{dq(r_{DA}, \Delta \varepsilon^0)}{dt} \sim k_0 |\exp(-0.5 \beta r_{DA})|^2 \exp\left(-\frac{[-\Delta \varepsilon^0 + \lambda]^2}{4 \lambda k_B T}\right) \qquad \text{eq. (1),}$$

for electrons tunneling through an enzyme's electrically-insulating peptide matrix between donor and acceptor sites (Figures 1a, b). The term $k_0$ gives a uni-molecular rate constant for electron transfer (~$10^{13}$ s$^{-1}$) at the minimum value of $r_{DA}$. Numerous changes to the tunneling decay constant, ($\beta$) and redox center reorganization energies ($\lambda$) may also be sampled over the course of protein evolution, marking additional, voltage-independent handles for ET rate control – and therefore catalytic control – in enzymes. Experimental confirmation of the Marcus description of charge transfer has been achieved through measurements of intramolecular electron transfer in various model systems, including *Pseudomonas aeruginosa* azurin[16–29], a variety of c-type cytochromes[16,30–32] and members of the P450 cytochrome family[33], highlighting how only variation in $r_{DA}$ permits the independent modulation of charge transfer rates, $dq(r_{DA}, \Delta\varepsilon)/dt$, from the electron potential, $\Delta\varepsilon$. This ability to independently control charge transfer rates and reaction driving force provides nature with an extra level of control over electron transfer kinetics and energetics. A contour of Marcus-dependent tunneling in Figure 1c summarizes this phenomenon, demonstrating how a range of tunneling currents spanning several orders of magnitude may be realized for a single value of the driving force (potential), through variation of $r_{DA}$. As a result, it becomes apparent that nature enjoys several degrees of freedom for modulating tunneling current densities separately from redox potentials in proteins. The distinction marks a potential advantage for catalytic tuning in PEC systems that is disallowed by the constraints of dark electrochemistry, where changes to reaction driving forces (applied potential) necessarily alter reaction kinetics (current).

As with the case of proteins, electrochemical decoupling may, in principle, be achieved by exploiting the distance-dependence of electron tunneling. Specifically, the establishment of thin, electrically-insulating layers between the electrode substrate and some catalyst layer of interest, with rate control at a given potential being granted by increasing or decreasing the insulator thickness (Figure 1b). Changes to the barrier thickness grant control over the electrode-catalyst (donor-acceptor) distance, and consequently, the electron tunneling current. This possibility has been borne-out in the field of molecular electronics, where previous studies on distance-dependent,

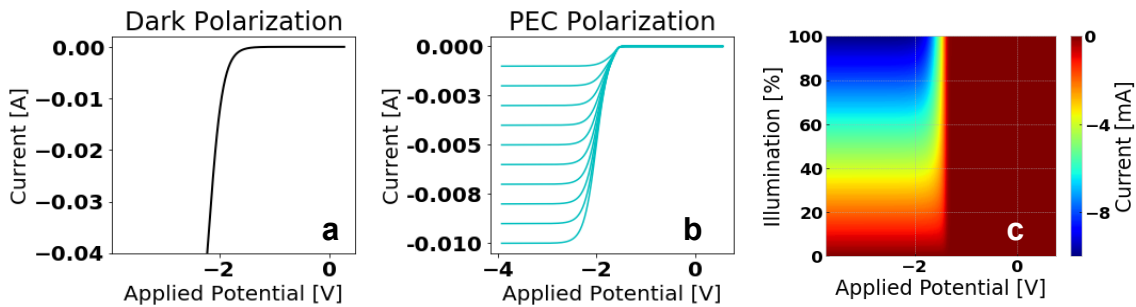

**Figure 2 (adapted from ref 36).**
**a.** Electrochemical polarization of a dark cell. Unlike with proteins, current and potential in dark electrochemical cells are coupled quantities, with testable [V,J] coordinates only being those lying on the catalyst polarization curve (black line). Calculated using eq. 1.

**b.** Simulated J-V response for photoelectrolyzer polarization. Here, independent control over electrocatalyst applied voltage and current is achieved by biasing the cell driving electrocatalysis. Current may be separately controlled by varying the intensity of light incident on the photoelectrolyzer. At high cell overpotentials, rates of electrocatalysis will be limited by the saturation current of the PV. With a photo-driven electrocatalyst, any polarization coordinate [V,J] in the region on or under the maximum current (-10 mA) polarization curve may now be tested. The result is a system where current and cell potential are functionally decoupled, with the set of accessible [V,J] coordinates now defined by the polarization curve integral at maximum illumination (filled cyan region). Light intensity may be varied continuously, resulting in accessible polarization states that are described by the integral of the photoelectrolyzer polarization curve at maximum illumination.

**c.** Contour mapping of this J-V decoupled electrocatalyst shows that current can be arbitrarily changed at a single voltage (by altering light illumination). Here, light acts as an additional degree of freedom for controlling current in analogy to the effects of distance-dependence in proteins, as shown in **Figure 1c**.

electron tunneling through insulating barriers, such as alkyl layers on Au electrodes and metal oxide films, have demonstrated the potential for controlling charge transfer to redox-active sites in this manner[15,34,35]. However, despite the significant body of work in this area, current-potential decoupling was not adopted as a general tool in the field of electrocatalysis. The omission is significant, as first-principles chemistry dictate that reactions will be under either thermodynamic or kinetic control. As a result, partitioned control between carrier kinetics and energetics may be reasonably expected to influence our control over product branching ratios in chemical reaction pathways.

**Current-potential decoupling in inorganic systems and descriptive chemical analogies**

Conventional dark voltammetry is conducted with an implicit understanding that current and potential cannot be independently controlled because the two mark co-dependent quantities, with variations in applied potential yielding corresponding changes in the current response of some given system of interest. The coupling between electrode current and potential is a relationship most commonly described using the Butler-Volmer formulation of interfacial electrode kinetics:

$$J(\eta) = J_0 \left[ \exp\left(\frac{F\eta\alpha}{RT}\right) - \exp\left(-\frac{F\eta(1-\alpha)}{RT}\right) \right] \qquad \text{eq. (2)},$$

where $\eta$ gives the electrode overpotential, $\alpha$ represents the symmetry factor, a term characterizing the transition state for electron transfer between donor and acceptor nuclear coordinates, and $J_0$ gives the equilibrium current density, defined at zero overpotential. Terms $F$, $R$ and $T$ represent Faraday's constant, the ideal gas gas constant, and system temperature, respectively. As a direct consequence of this, electrochemists treat the range of accessible operating conditions for an electrocatalyst as those combinations of current density and voltage (J-V coordinates, [V,J]) defining an electrocatalyst's polarization curve (Figure 2a). As a result, it is generally not possible to vary current without changing applied voltage (except in the specific cases where a polarization curve flattens).

While consideration of natural systems may be useful for illustrating the possibility of decoupling, attempting to tune current and applied potential independently, through step-wise changes in donor-acceptor distances in experimental systems, marks an inconvenient way of realizing decoupled electrocatalyst polarization. Such an approach would require that separate electrodes be fabricated for each donor-acceptor distance being tested. However, incorporation of light into electrochemical systems has been shown to make such explorations easier[36], with the independent control over current and voltage in the light-coupled apparatus yielding the same functional behavior as a system using distance-dependence as a mode of current-potential decoupling[36]. In general, the polarization response of semiconductor-liquid junctions incorporate aspects of both photovoltaics and dark electrocatalysts, with polarization curves that vary as both a function of potential and as a linear function of the light incident on the photoelectrode. Since incident light intensity may be varied by multiple means, such as placing a variable neutral density filter between the photoelectrode and light source, there exists a continuum of light-dependent, electrochemical response curves falling beneath the maximum-illumination polarization curve, that may now be accessed (Figure 2b). Here, our ability to vary electron energies by biasing the photoelectrode across a range of potentials (with a potentiostat), while independently limiting the total current – the rate of electron flow – by altering light intensity, yields a regime where current and applied potential are functionally decoupled. As a result, the expanded range of J-V coordinates in a decoupled system are now defined by the integral of the PEC polarization curve (with respect to applied potential) at maximum illumination. As a consequence, electrocatalyst polarization now assumes features akin to protein charge tunneling, with a continuum of current densities now accessible for a single value of applied potential/driving force (Figure 2c).

This expansion of possible polarization states is expected to have real implications for steering electrocatalyst selectivity. In the case of $CO_2$ reduction, the well-documented observation of electrochemical

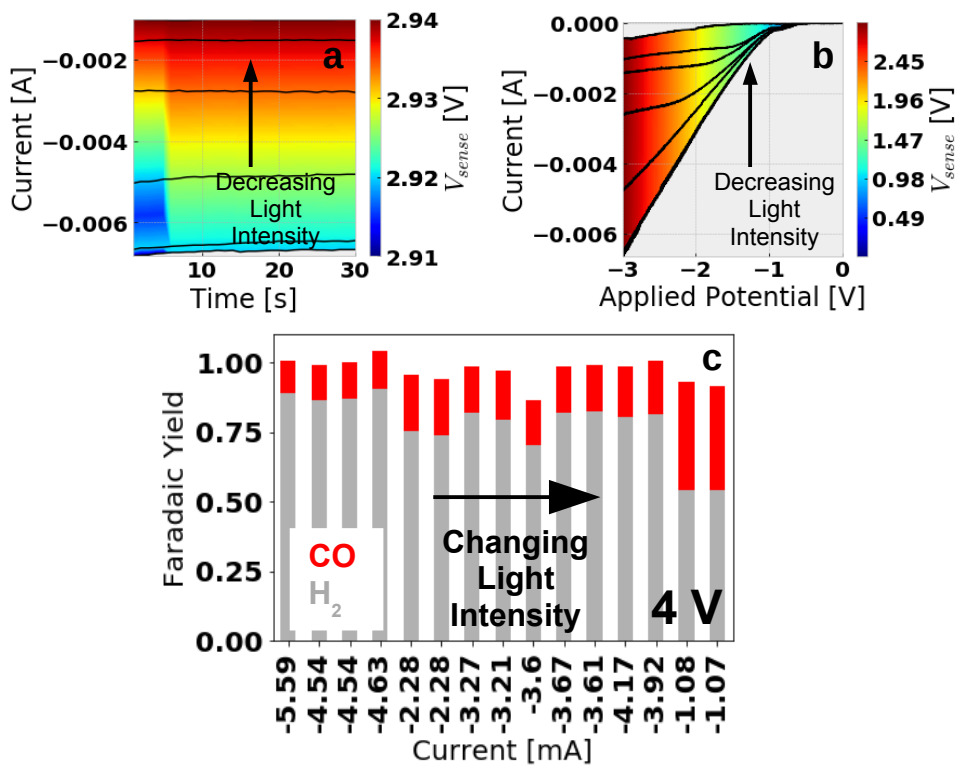

**Figure 3 (adapted from ref. 36)**
**a,b.** Monitoring cell potentials as light is used to change current density. Monitoring a floating voltage contact at the liquid-semiconductor interface (Vsense) demonstrates the possibilty of changing current magnitudes while stabilizing device operating potential, in a manner distinct from what is possible through PV-electrolyzer-type configurations.
**c.** Arbitrary current modulation in electrochemical $CO_2$ reduction reveals that product distributions are found to vary along both voltage and current axes, as opposed to just changing as a function of voltage.

hydrogen evolution (HER) kinetically out-competing $CO_2$ conversion at the high overpotentials where $CO_2$ reduction products also become energetically accessible[37], points to a useful role for decoupled polarization in mitigating adventitious HER. In particular, operating cathode catalysts in a low-current (kinetically limited), high overpotential (high energy) regime may enable $CO_2$ reduction with higher product selectivity. Notably, this option of selecting a cell/electrocatalyst polarization state of high overpotential and low current is generally precluded by dark voltammetry, where current scales solely as the exponent of applied potential. First-principles electrochemistry would suggest that operating electrocatalysts at decoupled current-potential coordinates lying beyond the dark polarization region (specifically in the low-current, high overpotential regime), may enable greater selectivity of $CO_2$RR over HER, while narrowing the distribution of reduced carbon products. Nascent efforts exploring decoupling in $CO_2$ conversion to syngas suggest this is indeed the case (Figure 3)[36], bolstering arguments for continuing these investigations on PEC decoupling in more complex catalytic systems. Explorations of Cu $CO_2$ reduction cathodes may prove especially illuminating, where the wide distribution of possible reduced-carbon products yielded during catalysis make matters of catalyst specificity and selectivity enhancements even more relevant.

    Appreciating the possibility of arbitrarily controlling current and applied potentials is most easily achieved by remembering that current density represents a flux, and as such, is influenced by more than a single parameter. Visualizing a stream of charge transiting through a conductive lattice at some potential $\varepsilon$, the energies associated

with any of these charges ($q\varepsilon$), will be attenuated through inelastic collisions of conduction band electrons with the atoms comprising the conductor. The kinetic energy of these charges are dissipated in such collisions as heat – the genesis of the resistive properties giving rise to $iR$ losses in an electrical circuit. The reduced kinetic energy of these electrons, as their mean-free path is dampened by lattice collisions, reduces their average speed – and therefore their time-averaged flux through the conductor. This ensemble charge flux is the measured current density, $J$, following the relation:

$$J(\epsilon) = nq\, v(\epsilon) \qquad \text{eq. (3a)}$$

where $n$ is the density of carriers (m$^{-3}$), $q$ gives the elementary charge (in C), and $v(\varepsilon)$ is the average electron velocity (m s$^{-1}$). Explicit description of the electron velocity's potential dependence is found by equating the potential energy of a charge in a conductor with a relativistic description of its kinetic energy, and solving for the particle velocity:

$$v(\epsilon) = \sqrt{\frac{q(\epsilon - iR) - m_e c^2}{2 m_e}} \qquad \text{eq. (3b).}$$

Here, $m_e$ gives the electron mass, $c$ is the speed of light, $R$ is the conductor resistance and $i$ is the current magnitude.

As implied by equation 3a, we can also imagine manipulating this flux by simply changing $n$ rather than $v$. In the case of metals, the significant occupancy of conduction band states at room temperature yields a high carrier concentration that remains essentially constant under ambient conditions, making it impossible to modulate $J$ through changes in $n$, forming the basis for why electrochemical polarization at metal and carbon electrodes is facilitated through applied potential modulation alone. However, semiconducting materials are unique in that their carrier concentrations are responsive to incident light absorption, opening a route for manipulating carrier fluxes through changes in carrier concentration, rather than the potential-dependent parameter $v$, as shown by equation 4a:

$$n \propto \gamma - k_{defect} - k_{rad} - k_{Auger} \qquad \text{eq. (4a),}$$

where $\gamma$ is the carrier generation rate, while $k_{defect}$, $k_{rad}$ and $k_{Auger}$ refer to charge recombination rates due to defects, radiative recombination, and Auger recombination, respectively. The light dependence of $n$ in a doped semiconductor arises from $\gamma$:

$$\gamma = \int_{x_1}^{x_2} \int_{\lambda_1}^{\lambda_2} I_0(\lambda)\, \alpha(\lambda)\, e^{-x\alpha(\lambda)}\, d\lambda\, dx \qquad \text{eq. (4b),}$$

where $I_0$ gives the spectrum-dependent incident photon flux, $\alpha(\lambda)$ gives the light absorption coefficient as a function of wavelength ($\lambda$), and $x$ is the depth from the illuminated semiconductor surface. Integration across all incident wavelengths, at all depths in the semiconductor from the illuminated surface, give the total carrier generation rate. As a result, light absorption can be seen to provide a secondary mode for controlling carrier flux – by altering $n$ rather than $v$ – in a manner that displays no explicit potential dependence.

This proposition of controlling carrier fluxes by manipulating carrier concentrations – rather than by parameters (such as potential) that influence actual mean carrier velocities – should not be too foreign to a chemist, largely because it is, in some sense, a restatement of effects already well-understood in physical chemistry. In electrochemistry, reaction fluxes under mass transport control represent the conflation of both diffusive rate constants of a species and the species concentration, as captured by the Fickian laws[38,39]. Furthermore, it is also well understood just how pronounced the effects of mass transport may be on steering the course of electrochemical reactivity[40]. As a result, it should be clear that in the case of mass-transport controlled reactions, currents – electron flows – may be regulated by controlling substrate delivery to an active site by changing either the substrate concentration or parameters that impact rates of diffusive transport (by changing electrolyte viscosity, for example), neither of which have voltage dependencies. The net effects are outcomes with close physical relationship to those posited by current-potential decoupling – specifically, using additional handles (here, changes to substrate mixing, diffusion rates or substrate concentration) as methods for enabling arbitrary current selection for a given applied potential. This is a point underscored by considering the qualitative similarity between polarization curves generated through Levich analysis using a rotating disk electrode and the light-limited curves generated using a photoelectrochemical device to independently vary current and applied potential. The latter instance, however, represents a way to achieve such functionality generally, for any reaction of choice, rather than the particular case of mass-transport-limited systems.

**Unifying decoupled protein electron transfer and artificial electrocatalysis**

The option of independently controlling carrier kinetics and carrier energetics should prompt a closer consideration of how redox-active proteins regulate their high catalytic specificities, with an eye on parlaying the physical principles by which these biochemical systems operate, into the development of artificial catalysts. The observation that nature uses a remarkably small suite of redox cofactors to achieve a wide range of redox-dependent transformations becomes less surprising, when stopping to realize the various ways by which reactivities of such active sites may be tuned to accommodate a wide range of chemistries. As Marcus Theory suggests, this may be achieved, in part, by regulating inter-site electron transfers independently of the potential through which the charge carrier travels, by positioning these redox-active motifs at unique distances in different proteins. Indeed, work on cytochrome P450s have shown that regulation of electron transfer rates alone – without changes in inter-site redox potential – are enough to impact the observation of photochemically-generated redox intermediates[41–43] and the product distributions resulting from electrochemical enzyme turnover[33].

Given that the range of testable [V,J] coordinates is infinitely greater in the decoupled case than it is for a conventional dark electrochemical experiment, testing a broad range of polarization conditions for a given catalyst may quickly become a time-prohibitive exercise, with product accumulation and measurement taking anywhere from minutes to hours for every polarization coordinate being tested. As a result, the task of testing different polarization conditions in a decoupled system, depending on the catalyst or step-resolution of the experiment, may range from time-consuming to intractable. However, previous advancements in our ability to quantitatively describe charge tunneling in biological systems may provide a useful route for circumventing this issue. In dynamic systems such as complex macromolecules, realistic determination of charge transfer rates must account for the time-averaged fluctuations in distance between two redox sites in a dynamic protein structure that may be rapidly sampling different conformations. As a result, the overall rate term becomes a weighted sum expressing the electronic coupling strength ($H_{DA}$) of various paths available for charge transfer between donor and acceptor sites in these systems[11,13,44–47]. Here we find ourselves at an advantage, as computational solutions for handling these complexities have already been developed. In particular, the pioneering computational methods of Beratan and Onuchic have facilitated the path-dependent calculation of inter-site electron transfers in proteins with solved structures, permitting the explicit determination of otherwise hard-to-know electron transfer rates in proteins[46–52]. Beratan and Onuchic's approach – codified by Balabin in the *PATHWAYS* algorithm – solves the problem of how to determine possible competing routes for electron transfer by calculating the electron coupling between donor and acceptor for every possible charge transfer path. Overall couplings are calculated by decomposing transfer paths into a series of discrete steps comprised of charge hopping or tunneling through covalent bonds, hydrogen bonds and through-space jumps[52]:

$$H_{DA} \propto \prod_i w_i^c \prod_j w_j^h \prod_k w_k^s \qquad \text{eq. (5)}.$$

The method embodied in eq. 5 determines $H_{DA}$ of a given pathway, being expressed as the product of a series of individual covalent bond ($w^c$), hydrogen bond ($w^h$), and through-space ($w^s$) electronic couplings that occur between adjacent sites for electron hopping between an arbitrarily located donor and acceptor.

Such capabilities, paired with an ability to determine relevant redox potentials using established analytical methods such as spectroelectrochemical titration, provide chemists with the tools necessary for determining the effective current-potential coordinate at which a given enzyme operates. Application of such information to decoupled regulation of charge transfer could then help identify promising [V,J] coordinates for a related inorganic catalyst, where catalysis may be optimal. For example, it is conceivable that $MoS_2$ and related 2D materials may be competent for $CO_2$ reduction under aqueous media. However, efforts to date have only succeeded in using these materials in proton reduction[53–62] and $CO_2$ reduction in non-aqueous media[63,64]. This is despite the perfect fidelity with which molybdo-sulfur and tungsto-sulfur formate dehydrogenase enzymes interconvert $CO_2$ and formate. However, incorporation of decoupled polarization may provide us with an expanded set of conditions under which we can test such materials; the knowledge of the [V,J] operating points characterizing a redox enzyme may help us reduce the time needed to explore this expanded range of polarization conditions in a J-V decoupled, PEC device. Conversely, pursuing this line of scientific inquiry may also reveal fundamental relationships between current-potential coordinates characterizing a redox enzyme and the chemical transformations it can facilitate. The possibility of uncovering such correlations is suggested by recent findings, where utilization of decoupling revealed the existence of a smooth relation between [V,J] coordinate (rather than voltage alone as is commonly assumed) and product ratios in electrochemical $CO_2$ reduction[36].

**Outlook**

Independent manipulation of current and voltage as handles for steering electrochemical reactions should motivate a reexamination of the predominant approaches to biomimetic electrocatalysis. While biophysical design emphases and synthetic bioinorganic approaches are often spoken of interchangeably, the distinctions between the two are critical, and are perhaps best highlighted via analogy: both a bird and a helicopter can fly, though the physical resemblance between the two is weak. Early attempts to build a flying machine by assuming that it must look like a bird would have precluded the helicopter's invention. However, a general comprehension of Bernoulli's physical Principle of Lift opens up the possibility of using a narrow, rotary blade instead of a broad, feathered wing for flight. The comparison demonstrates how aesthetic dissimilarity between two structures may nevertheless yield a functional convergence, so long as they operate under the same physical principle. This highlights the key difference between mimicking natural catalytic schemes using biophysical vs synthetic bioinorganic approaches. In the latter case, efforts to replicate enzyme chemistries have placed an overwhelming emphasis on the assumed essence of structure-function correlations, a fact exemplified by the synthesis of molecular complexes that resemble protein active sites but rarely yield reactivities comparable to the corresponding enzyme. However, recent findings suggest that an aim of catalyst development should be to incorporate more of the biophysical design paradigm alongside synthetic bioinorganic treatments, through an increased focus on emulating the underlying physics of enzyme functioning, rather than operating under assumptions that structural similarities to protein active sites alone will be enough to get us there. In the case of reproducing nature's decoupled regulation of electron transport in enzyme catalysis, decoupling may provide an important step forward, as the field of artificial electrocatalysis steadily advances towards fully functional biomimicry.

**References**

(1) Kanady, J. S.; Tsui, E. Y.; Day, M. W.; Agapie, T. A Synthetic Model of the Mn3Ca Subsite of the Oxygen-Evolving Complex in Photosystem II. *Science* **2011**, *333* (6043), 733–736. https://doi.org/10.1126/science.1206036.

(2) Kanady, J. S.; Lin, P.-H.; Carsch, K. M.; Nielsen, R. J.; Takase, M. K.; Goddard, W. A.; Agapie, T. Toward Models for the Full Oxygen-Evolving Complex of Photosystem II by Ligand Coordination To Lower the Symmetry of the Mn3CaO4 Cubane: Demonstration That Electronic Effects Facilitate Binding of a Fifth Metal. *J Am Chem Soc* **2014**, *136* (41), 14373–14376. https://doi.org/10.1021/ja508160x.

(3) Hlavica, P. Models and Mechanisms of O-O Bond Activation by Cytochrome P450. *European Journal of Biochemistry* **2004**, *271* (22), 4335–4360. https://doi.org/10.1111/j.1432-1033.2004.04380.x.

(4) Kuriyan, J. *The Molecules of Life : Physical and Chemical Principles /*; Garland Science, Taylor & Francis Group,: New York :, c2013.

(5) Marcus, R. A.; Sutin, N. Electron Transfers in Chemistry and Biology. *Biochimica et Biophysica Acta (BBA) - Reviews on Bioenergetics* **1985**, *811* (3), 265–322. https://doi.org/10.1016/0304-4173(85)90014-X.

(6) *Water oxidation catalysis by manganese oxides: learning from evolution - Energy & Environmental Science (RSC Publishing)*. https://pubs.rsc.org/en/content/articlelanding/2014/ee/c4ee00681j#!divAbstract (accessed 2020-08-07).

(7) Umena, Y.; Kawakami, K.; Shen, J.-R.; Kamiya, N. Crystal Structure of Oxygen-Evolving Photosystem II at a Resolution of 1.9 Å. *Nature* **2011**, *473* (7345), 55–60. https://doi.org/10.1038/nature09913.

(8) Winkler, J. R.; Di Bilio, A.; Farrow, N. A.; Richards, J. H.; Gray, H. B. Electron Tunneling in Biological Molecules. *Pure and Applied Chemistry* **1999**, *71* (9), 1753–1764. https://doi.org/10.1351/pac199971091753.

(9) Gray, H. B.; Winkler, J. R. Long-Range Electron Transfer. *PNAS* **2005**, *102* (10), 3534–3539. https://doi.org/10.1073/pnas.0408029102.

(10) Gray, H. B.; Winkler, J. R. Electron Tunneling Through Proteins. *Quarterly Reviews of Biophysics* **2003**, *36* (03), 341–372. https://doi.org/10.1017/S0033583503003913.

(11) Warren, J. J.; Ener, M. E.; Vlček Jr., A.; Winkler, J. R.; Gray, H. B. Electron Hopping through Proteins. *Coordination Chemistry Reviews* No. 0. https://doi.org/10.1016/j.ccr.2012.03.032.

(12) Saen-Oon, S.; Lucas, M. F.; Guallar, V. Electron Transfer in Proteins: Theory, Applications and Future Perspectives. *Phys. Chem. Chem. Phys.* **2013**. https://doi.org/10.1039/C3CP50484K.

(13) Voityuk, A. A. Long-Range Electron Transfer in Biomolecules. Tunneling or Hopping? *J. Phys. Chem. B* **2011**, *115* (42), 12202–12207. https://doi.org/10.1021/jp2054876.

(14) Finklea, H. O.; Hanshew, D. D. Electron-Transfer Kinetics in Organized Thiol Monolayers with Attached Pentaammine(Pyridine)Ruthenium Redox Centers. *J. Am. Chem. Soc.* **1992**, *114* (9), 3173–3181. https://doi.org/10.1021/ja00035a001.

(15) Smalley, J. F.; Feldberg, S. W.; Chidsey, C. E. D.; Linford, M. R.; Newton, M. D.; Liu, Y.-P. The Kinetics of Electron Transfer Through Ferrocene-Terminated Alkanethiol Monolayers on Gold. *J. Phys. Chem.* **1995**, *99* (35), 13141–13149. https://doi.org/10.1021/j100035a016.

(16) Cummins, D.; Gray, H. B. Electron-Transfer Protein Reactivities. Kinetic Studies of the Oxidation of Horse Heart Cytochrome c, Chromatium Vinosum High Potential Iron-Sulfur Protein, Pseudomonas Aeruginosa Azurin, Bean Plastocyanin, and Rhus Vernicifera Stellacyanin by Pentaammminepyridineruthenium (III). *Journal of the American Chemical Society* **1977**, *99* (15), 5158–5167.

(17) Bordi, F.; Prato, M.; Cavalleri, O.; Cametti, C.; Canepa, M.; Gliozzi, A. Azurin Self-Assembled Monolayers Characterized by Coupling Electrical Impedance Spectroscopy and Spectroscopic Ellipsometry. *The Journal of Physical Chemistry B* **2004**, *108* (52), 20263–20272. https://doi.org/10.1021/jp047141y.

(18) Farver, O.; Skov, L. K.; Young, S.; Bonander, N.; Karlsson, B. G.; Vänng\aard, T.; Pecht, I. Aromatic

(19) Lancaster, K. M.; Farver, O.; Wherland, S.; Crane, E. J.; Richards, J. H.; Pecht, I.; Gray, H. B. Electron Transfer Reactivity of Type Zero Pseudomonas Aeruginosa Azurin. *J. Am. Chem. Soc.* **2011**, *133* (13), 4865–4873. https://doi.org/10.1021/ja1093919.

(20) Farver, O.; Pecht, I. Long Range Intramolecular Electron Transfer in Azurins. *J. Am. Chem. Soc.* **1992**, *114* (14), 5764–5767. https://doi.org/10.1021/ja00040a043.

(21) Farver, O.; Lu, Y.; Ang, M. C.; Pecht, I. Enhanced Rate of Intramolecular Electron Transfer in an Engineered Purple CuA Azurin. *PNAS* **1999**, *96* (3), 899–902. https://doi.org/10.1073/pnas.96.3.899.

(22) Gray, H. B.; Winkler, J. R. Electron Flow through Metalloproteins. *Biochimica et Biophysica Acta (BBA) - Bioenergetics* **2010**, *1797* (9), 1563–1572. https://doi.org/10.1016/j.bbabio.2010.05.001.

(23) Yokoyama, K.; Leigh, B. S.; Sheng, Y.; Niki, K.; Nakamura, N.; Ohno, H.; Winkler, J. R.; Gray, H. B.; Richards, J. H. Electron Tunneling through Pseudomonas Aeruginosa Azurins on SAM Gold Electrodes. *Inorganica Chim Acta* **2008**, *361* (4), 1095–1099. https://doi.org/10.1016/j.ica.2007.08.022.

(24) Fujita, K.; Nakamura, N.; Ohno, H.; Leigh, B. S.; Niki, K.; Gray, H. B.; Richards, J. H. Mimicking Protein−Protein Electron Transfer:  Voltammetry of Pseudomonas Aeruginosa Azurin and the Thermus Thermophilus CuA Domain at ω-Derivatized Self-Assembled-Monolayer Gold Electrodes. *J. Am. Chem. Soc.* **2004**, *126* (43), 13954–13961. https://doi.org/10.1021/ja047875o.

(25) Shih, C.; Museth, A. K.; Abrahamsson, M.; Blanco-Rodriguez, A. M.; Bilio, A. J. D.; Sudhamsu, J.; Crane, B. R.; Ronayne, K. L.; Towrie, M.; Vlček, A.; Richards, J. H.; Winkler, J. R.; Gray, H. B. Tryptophan-Accelerated Electron Flow Through Proteins. *Science* **2008**, *320* (5884), 1760–1762. https://doi.org/10.1126/science.1158241.

(26) Sepunaru, L.; Pecht, I.; Sheves, M.; Cahen, D. Solid-State Electron Transport across Azurin: From a Temperature-Independent to a Temperature-Activated Mechanism. *J. Am. Chem. Soc.* **2011**, *133* (8), 2421–2423. https://doi.org/10.1021/ja109989f.

(27) Farver, O.; Marshall, N. M.; Wherland, S.; Lu, Y.; Pecht, I. Designed Azurins Show Lower Reorganization Free Energies for Intraprotein Electron Transfer. *PNAS* **2013**, *110* (26), 10536–10540. https://doi.org/10.1073/pnas.1215081110.

(28) Crane, B. R.; Di Bilio, A. J.; Winkler, J. R.; Gray, H. B. Electron Tunneling in Single Crystals of Pseudomonas Aeruginosa Azurins. *Journal of the American Chemical Society* **2001**, *123* (47), 11623–11631. https://doi.org/10.1021/ja0115870.

(29) Khoshtariya, D. E.; Dolidze, T. D.; Shushanyan, M.; Davis, K. L.; Waldeck, D. H.; Eldik, R. van. Fundamental Signatures of Short- and Long-Range Electron Transfer for the Blue Copper Protein Azurin at Au/SAM Junctions. *PNAS* **2010**, *107* (7), 2757–2762. https://doi.org/10.1073/pnas.0910837107.

(30) Gray, H. B.; Winkler, J. R. Electron Flow through Proteins. *Chemical Physics Letters* **2009**, *483* (1–3), 1–9. https://doi.org/10.1016/j.cplett.2009.10.051.

(31) Kuznetsov, B. A.; Byzova, N. A.; Shumakovich, G. P. The Effect of the Orientation of Cytochrome c Molecules Covalently Attached to the Electrode Surface upon Their Electrochemical Activity. *Journal of Electroanalytical Chemistry* **1994**, *371* (1–2), 85–92. https://doi.org/10.1016/0022-0728(93)03219-F.

(32) Alvarez-Paggi, D.; Meister, W.; Kuhlmann, U.; Weidinger, I.; Tenger, K.; Zimányi, L.; Rákhely, G.; Hildebrandt, P.; Murgida, D. H. Disentangling Electron Tunneling and Protein Dynamics of Cytochrome c through a Rationally Designed Surface Mutation. *J. Phys. Chem. B* **2013**. https://doi.org/10.1021/jp400832m.

(33) van der Felt, C.; Hindoyan, K.; Choi, K.; Javdan, N.; Goldman, P.; Bustos, R.; Star, A. G.; Hunter, B. M.; Hill, M. G.; Nersissian, A.; Udit, A. K. Electron-Transfer Rates Govern Product Distribution in Electrochemically-Driven P450-Catalyzed Dioxygen Reduction. *Journal of Inorganic Biochemistry* **2011**,

Residues May Enhance Intramolecular Electron Transfer in Azurin. *Journal of the American Chemical Society* **1997**, *119* (23), 5453–5454. https://doi.org/10.1021/ja964386i.

*105* (10), 1350–1353. https://doi.org/10.1016/j.jinorgbio.2011.03.006.

(34) Devaraj, N. K.; Decreau, R. A.; Ebina, W.; Collman, J. P.; Chidsey, C. E. D. Rate of Interfacial Electron Transfer through the 1,2,3-Triazole Linkage. *J. Phys. Chem. B* **2006**, *110* (32), 15955–15962. https://doi.org/10.1021/jp057416p.

(35) Neuhausen, A. B.; Hosseini, A.; Sulpizio, J. A.; Chidsey, C. E. D.; Goldhaber-Gordon, D. Molecular Junctions of Self-Assembled Monolayers with Conducting Polymer Contacts. *ACS Nano* **2012**, *6* (11), 9920–9931. https://doi.org/10.1021/nn3035183.

(36) Agbo, P. *An Expansion of Polarization Control Using Semiconductor-Liquid Contacts*. arXiv.org. https://arxiv.org/abs/2306.00143v1.

(37) Huan, T. N.; Corte, D. A. D.; Lamaison, S.; Karapinar, D.; Lutz, L.; Menguy, N.; Foldyna, M.; Turren-Cruz, S.-H.; Hagfeldt, A.; Bella, F.; Fontecave, M.; Mougel, V. Low-Cost High-Efficiency System for Solar-Driven Conversion of CO2 to Hydrocarbons. *PNAS* **2019**, *116* (20), 9735–9740. https://doi.org/10.1073/pnas.1815412116.

(38) Bockris, J. O.; Reddy, A. K. N.; Gamboa-Aldeco, M. *Modern Electrochemistry*; Springer, 2000.

(39) Bard, A. J.; Faulkner, L. R. *Electrochemical Methods: Fundamentals and Applications*; Wiley, 2001.

(40) Watkins, N. B.; Schiffer, Z. J.; Lai, Y.; Musgrave, C. B. I.; Atwater, H. A.; Goddard, W. A. I.; Agapie, T.; Peters, J. C.; Gregoire, J. M. Hydrodynamics Change Tafel Slopes in Electrochemical CO2 Reduction on Copper. *ACS Energy Lett.* **2023**, *8* (5), 2185–2192. https://doi.org/10.1021/acsenergylett.3c00442.

(41) Ener, M. E.; Gray, H. B.; Winkler, J. R. Hole Hopping through Tryptophan in Cytochrome P450. *Biochemistry* **2017**, *56* (28), 3531–3538. https://doi.org/10.1021/acs.biochem.7b00432.

(42) Krest, C. M.; Onderko, E. L.; Yosca, T. H.; Calixto, J. C.; Karp, R. F.; Livada, J.; Rittle, J.; Green, M. T. Reactive Intermediates in Cytochrome P450 Catalysis*. *Journal of Biological Chemistry* **2013**, *288* (24), 17074–17081. https://doi.org/10.1074/jbc.R113.473108.

(43) Ener, M. E.; Lee, Y.-T.; Winkler, J. R.; Gray, H. B.; Cheruzel, L. Photooxidation of Cytochrome P450-BM3. *Proceedings of the National Academy of Sciences* **2010**, *107* (44), 18783–18786. https://doi.org/10.1073/pnas.1012381107.

(44) Ricks, A. B.; Solomon, G. C.; Colvin, M. T.; Scott, A. M.; Chen, K.; Ratner, M. A.; Wasielewski, M. R. Controlling Electron Transfer in Donor−Bridge−Acceptor Molecules Using Cross-Conjugated Bridges. *J. Am. Chem. Soc.* **2010**, *132* (43), 15427–15434. https://doi.org/10.1021/ja107420a.

(45) Warren, J. J.; Winkler, J. R.; Gray, H. B. Hopping Maps for Photosynthetic Reaction Centers. *Coordination Chemistry Reviews* No. 0. https://doi.org/10.1016/j.ccr.2012.07.002.

(46) Walther, M. E.; Wenger, O. S. Hole Tunneling and Hopping in a Ru(Bpy)32+-Phenothiazine Dyad with a Bridge Derived from Oligo-p-Phenylene. *Inorg. Chem.* **2011**, *50* (21), 10901–10907. https://doi.org/10.1021/ic201446x.

(47) Wenger, O. S. How Donor−Bridge−Acceptor Energetics Influence Electron Tunneling Dynamics and Their Distance Dependences. *Acc. Chem. Res.* **2010**, *44* (1), 25–35. https://doi.org/10.1021/ar100092v.

(48) Beratan, D. N.; Betts, J. N.; Onuchic, J. N. Tunneling Pathway and Redox-State-Dependent Electronic Couplings at Nearly Fixed Distance in Electron Transfer Proteins. *J. Phys. Chem.* **1992**, *96* (7), 2852–2855. https://doi.org/10.1021/j100186a014.

(49) Beratan, D. N.; Betts, J. N.; Onuchic, J. N. Protein Electron Transfer Rates Set by the Bridging Secondary and Tertiary Structure. *Science* **1991**, *252* (5010), 1285–1288. https://doi.org/10.1126/science.1656523.

(50) Balabin, I. A.; Beratan, D. N.; Skourtis, S. S. Persistence of Structure Over Fluctuations in Biological Electron-Transfer Reactions. *Phys. Rev. Lett.* **2008**, *101* (15), 158102. https://doi.org/10.1103/PhysRevLett.101.158102.

(51) Onuchic, J. N.; Beratan, D. N.; Winkler, J. R.; Gray, H. B. Pathway Analysis of Protein Electron-Transfer Reactions. *Annual Review of Biophysics and Biomolecular Structure* **1992**, *21* (1), 349–377. https://doi.org/

(52) Balabin, I. A.; Hu, X.; Beratan, D. N. Exploring Biological Electron Transfer Pathway Dynamics with the Pathways Plugin for VMD. *J Comput Chem* **2012**, *33* (8), 906–910. https://doi.org/10.1002/jcc.22927.

(53) Lukowski, M. A.; Daniel, A. S.; Meng, F.; Forticaux, A.; Li, L.; Jin, S. Enhanced Hydrogen Evolution Catalysis from Chemically Exfoliated Metallic MoS2 Nanosheets. *J. Am. Chem. Soc.* **2013**. https://doi.org/10.1021/ja404523s.

(54) Laursen, A. B.; Pedersen, T.; Malacrida, P.; Seger, B.; Hansen, O.; Vesborg, P. C. K.; Chorkendorff, I. MoS2—an Integrated Protective and Active Layer on N+p-Si for Solar H2 Evolution. *Phys. Chem. Chem. Phys.* **2013**, *15* (46), 20000–20004. https://doi.org/10.1039/C3CP52890A.

(55) Kong, D.; Wang, H.; Cha, J. J.; Pasta, M.; Koski, K. J.; Yao, J.; Cui, Y. Synthesis of MoS2 and MoSe2 Films with Vertically Aligned Layers. *Nano Lett.* **2013**, *13* (3), 1341–1347. https://doi.org/10.1021/nl400258t.

(56) Xue, N.; Diao, P. Composite of Few-Layered MoS2 Grown on Carbon Black: Tuning the Ratio of Terminal to Total Sulfur in MoS2 for Hydrogen Evolution Reaction. *J. Phys. Chem. C* **2017**, *121* (27), 14413–14425. https://doi.org/10.1021/acs.jpcc.7b02522.

(57) *Electrochemical generation of sulfur vacancies in the basal plane of MoS2 for hydrogen evolution - ncomms15113.pdf*. https://www.nature.com/articles/ncomms15113.pdf (accessed 2018-01-14).

(58) Chang, K.; Chen, W. In Situ Synthesis of MoS2/Graphene Nanosheet Composites with Extraordinarily High Electrochemical Performance for Lithium Ion Batteries. *Chem. Commun.* **2011**, *47* (14), 4252–4254. https://doi.org/10.1039/C1CC10631G.

(59) King, L. A.; Hellstern, T. R.; Park, J.; Sinclair, R.; Jaramillo, T. F. Highly Stable Molybdenum Disulfide Protected Silicon Photocathodes for Photoelectrochemical Water Splitting. *ACS Appl. Mater. Interfaces* **2017**, *9* (42), 36792–36798. https://doi.org/10.1021/acsami.7b10749.

(60) Xue, N.; Diao, P. Composite of Few-Layered MoS2 Grown on Carbon Black: Tuning the Ratio of Terminal to Total Sulfur in MoS2 for Hydrogen Evolution Reaction. *J. Phys. Chem. C* **2017**, *121* (27), 14413–14425. https://doi.org/10.1021/acs.jpcc.7b02522.

(61) Jaramillo, T. F.; Jorgensen, K. P.; Bonde, J.; Nielsen, J. H.; Horch, S.; Chorkendorff, I. Identification of Active Edge Sites for Electrochemical H2 Evolution from MoS2 Nanocatalysts. *Science* **2007**, *317* (5834), 100–102. https://doi.org/10.1126/science.1141483.

(62) Wu, Z.; Fang, B.; Wang, Z.; Wang, C.; Liu, Z.; Liu, F.; Wang, W.; Alfantazi, A.; Wang, D.; Wilkinson, D. P. MoS2 Nanosheets: A Designed Structure with High Active Site Density for the Hydrogen Evolution Reaction. *ACS Catal.* **2013**, *3* (9), 2101–2107. https://doi.org/10.1021/cs400384h.

(63) Abbasi, P.; Asadi, M.; Liu, C.; Sharifi-Asl, S.; Sayahpour, B.; Behranginia, A.; Zapol, P.; Shahbazian-Yassar, R.; Curtiss, L. A.; Salehi-Khojin, A. Tailoring the Edge Structure of Molybdenum Disulfide toward Electrocatalytic Reduction of Carbon Dioxide. *ACS Nano* **2017**, *11* (1), 453–460. https://doi.org/10.1021/acsnano.6b06392.

(64) Asadi, M.; Kim, K.; Liu, C.; Addepalli, A. V.; Abbasi, P.; Yasaei, P.; Phillips, P.; Behranginia, A.; Cerrato, J. M.; Haasch, R.; Zapol, P.; Kumar, B.; Klie, R. F.; Abiade, J.; Curtiss, L. A.; Salehi-Khojin, A. Nanostructured Transition Metal Dichalcogenide Electrocatalysts for CO2 Reduction in Ionic Liquid. *Science* **2016**, *353* (6298), 467–470. https://doi.org/10.1126/science.aaf4767.

**Acknowledgements**

This work was supported by the Liquid Sunlight Alliance a DOE Energy Innovation Hub, supported through the Office of Science of the U.S. Department of Energy under Award Number DE-SC0004993.

**Competing Interests**